\documentclass[twocolumn, prl,preprintnumbers,amsmath,amssymb]{revtex4}
\usepackage{amsmath,dcolumn,graphicx,epsf,ulem,subfigure, array}
\setkeys{Gin}{width=8.6cm,keepaspectratio}
\usepackage{dcolumn}
\usepackage{bm}

\begin{document}

\title{All-Inorganic Spin-Cast Nanoparticle Solar Cells with Non-Selective Electrodes}
\author{{\bf I. E. Anderson$^1$, A.J. Breeze$^2$, J.D. Olson, L. Yang, Y. Sahoo$^3$, S.A. Carter$^1$}}
\affiliation{$^1$Department of Physics, University of California, Santa Cruz, California 95064, USA}
\affiliation{$^2$Solexant Corp., San Jose, California 95131, USA}
\affiliation{$^3$Institute for Lasers, Photonics and Biophotonics, Department of Chemistry, The State University of New York at Buffalo, Buffalo, New York 14260-4200, USA}

\date{\today}

\begin{abstract}{{\bf Abstract:}Spin-cast all-inorganic nanoparticle solutions have been used to make a CdTe/CdSe solar cell with an efficiency of up to 2.6\% without alumina or calcium buffer layers. The type of junction as well as the non-selective nature of the electrodes of these devices are explored.}
\end{abstract}

\maketitle

Semiconductor nanoparticles exhibit many unique properties which have implications for solar cells. Their bandgaps are easily tuned by changing nanoparticle size\cite {ee82}. In addition, they gain some practical properties of organic materials, such as solution processibility, while retaining some useful properties of inorganic semiconductors, such as thermal and electrical stability. Finally, nanoparticles may also decrease the probability of non-radiative decay to phonons, allowing for the generation of multiple excitons through impact ionization\cite {sk04,ebj05}.

Previous results of Gur, et al.\cite {gfg05}, suggest that all-inorganic nanoparticle solar cells consisting of CdTe and CdSe nanorods act as type-II heterojunctions, and require a thin layer of alumina and a top electrode of Ca for high efficiency. We have made a slightly thicker device that eliminates the need for the alumina and Ca layers, and provide some explanation for the physical nature of the junction and non-selective electrodes.

\begin{figure}[b]
\includegraphics[width=3.25in,totalheight=2.5in,angle=0]{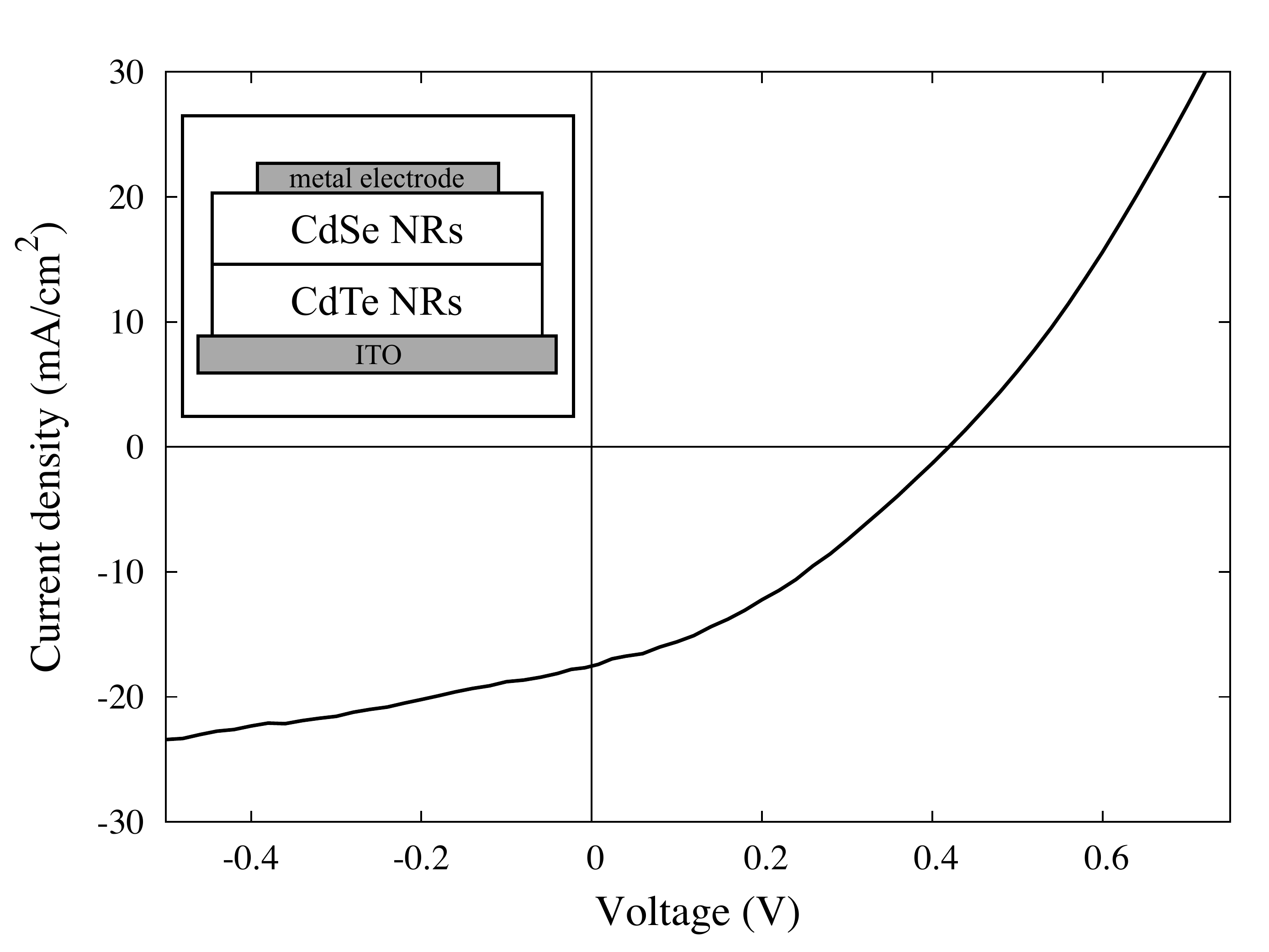}
\caption{J-V curve for a device under AM1.5G illumination. $\rm J_{SC}$ is 16.9 $\rm mA/cm^3$, $\rm V_{OC}$ is 0.42V, and fill factor is 36\%; total efficiency is 2.6\%. Device structure consists of layers of nanoparticles (inset).}
\label{structure}
\end{figure}

We made solar cells with rod-shaped $\rm CdTe$ and $\rm CdSe$ in pyridine provided by Solexant Corp. Devices were fabricated by spin coating $\rm CdTe$ followed by $\rm CdSe$ on a patterned indium tin oxide (ITO) substrate and annealing at 215-220C between layers. The device was treated with $\rm CdCl_2$ in methanol and then sintered in air at 400C. Light curves were taken with simulated Air Mass 1.5 Global (AM1.5G) illumination from the ITO side. Fig. \ref {structure} shows the structure of these devices, as well as the J-V characteristics of the cells. The most efficient device in this study has a CdTe layer of 180 nm and a CdSe layer of 70 nm.

\begin{figure}
\includegraphics[width=3.25in,totalheight=3in,angle=0]{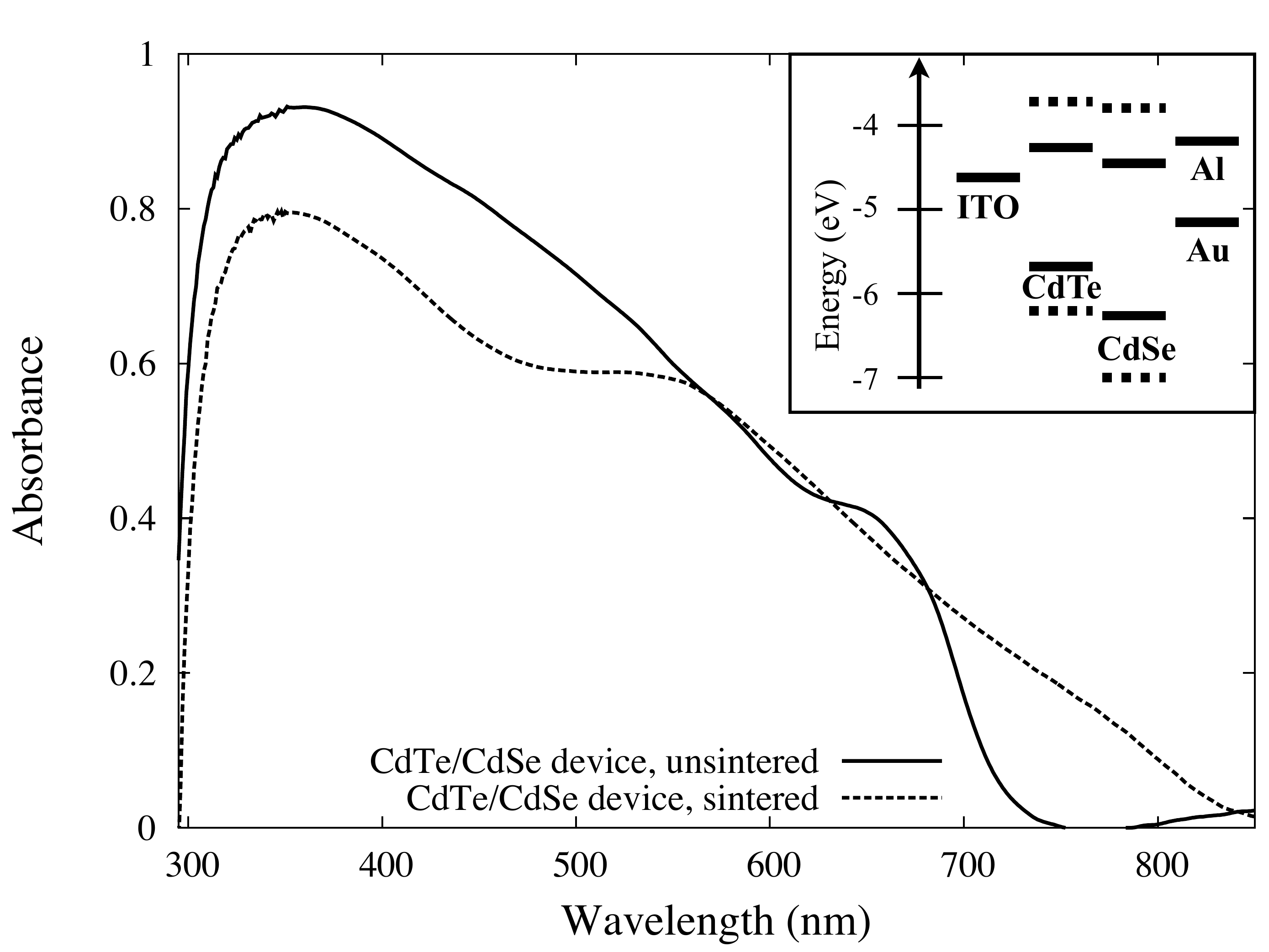}
\caption{Absorption data before and after sintering. Before sintering, the absorption is blue-shifted, signifying a bigger bandgap and quantum confinement. {\bf (inset)}Band structure of bulk CdTe and CdSe, as well as ITO and Au or Al electrodes. Solid lines are for bulk band edges, while dotted lines denote band edges calculated with an effective mass approximation assuming quantum confinement and parabolic bands.}
\label{absorb}
\end{figure}

An estimate of the bandgap with and without quantum confinement is shown in Fig. \ref {absorb}; the calculation is done with an effective mass approximation for a 1D system\cite {yll03}, although these calculations are dependent on nanoparticle shape distribution\cite {erk96}, so the energy levels should be taken as approximate. Absorbance measurements (Fig. \ref {absorb}) show that photons are still being absorbed at 860 nm in the sintered devices. This corresponds to CdTe's  bulk bandgap of 1.44eV; therefore, sintering appears to recover the bulk behavior of CdTe and CdSe. Absorbance measurements for films of CdSe confirm that its bandgap also diminishes to the bulk value after sintering. Before sintering, the devices have cutoff wavelengths similar to (within 10 nm of) CdTe films that have never been heated. When we made devices without sintering, they had low currents of about 100$\mu$A, presumably because charge cannot easily pass between particles.

The prior report on CdTe/CdSe nanoparticle solar cells claims both CdTe and CdSe are required to make a working solar cell. However, we are able to synthesize a solar cell similar to Fig. \ref {structure} but without the CdSe layer. The so-called CdTe-only solar cell's active layer is about 400 nm thick. Prior CdTe-only devices were 200 nm thick, and it may be that these devices were hindered by shunting paths through the thinner CdTe film, or that, as in thin-film CdS/CdTe devices\cite {ccf91, bf93}, the region of optical absorption into the CdTe was too close to a surface and recombination sites decreased the devices' efficiency.

In the work done by Gur, et. al, the inability of the CdTe- or CdSe-only devices to work, along with the large sheet resistance of both the CdTe and CdSe films, was provided as proof for a type-II heterojunction. The true sheet resistance of these films should not be measured on glass, but rather after exposure to ITO. This is impossible to do, as the current would likely pass perpendicular to the film and ultimately through the ITO electrode. We cannot therefore rule out the possibility that the CdTe is doped by the ITO; in fact, early work on CdTe p-n cells utilized indium as an n-type dopant\cite{r59}.

The CdTe-only cell is not a heterojunction because there is no other material with which to form the necessary band offsets. All J-V measurements for both the CdTe and CdTe/CdSe cells show Al collecting electrons and ITO collecting holes. If indium doped the CdTe near the ITO, and a p-n junction were formed, charge would flow in the opposite direction. We therefore conclude that dissociation of excitons in the CdTe-only device must be performed by a Schottky barrier. The aluminum forms an ohmic contact with the CdTe, and most Schottky barrier solar cells are designed so that the light penetrates the side of the active junction\cite{g75}; this is especially pertinent for CdTe because of its short absorption length. The CdTe-Al contact is therefore not a  good candidate for dissociating excitons. However, in order for a Schottky barrier at the CdTe-ITO interface to function, holes must be collected at this interface, and the majority carriers must be electrons. This suggests that the In doping plays a crucial role in barrier formation by n-doping the CdTe. Without an insulating layer the Schottky barrier solar cell's $\rm V_{oc}$ has historically been limited to 600 mV\cite{cy76}, which is consistent with our data.

\begin{figure}
\includegraphics[width=3.25in,angle=0]{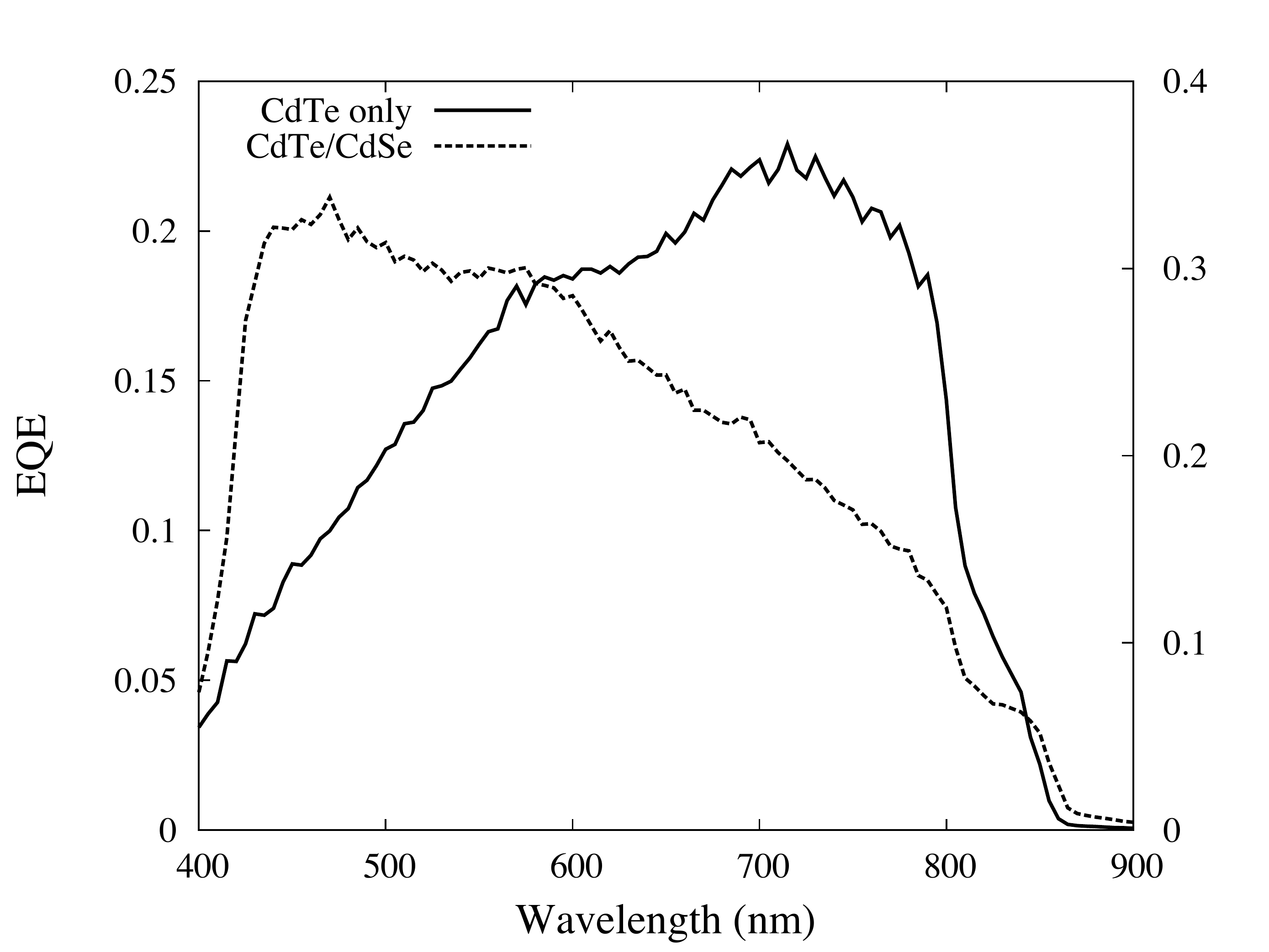}
\caption{External quantum efficiency data corresponding to the J-V curves in Fig. \ref{IV2}. The CdTe/CdSe device is plotted on the left axis, while the CdTe-only device is plotted on the right. The difference in intensity of the two spectra is different because the CdTe/CdSe device was old and had been exposed to oxygen before the EQE data was taken. The data is meant to show the relative spectral response.}
\label{EQE}
\end{figure}

\begin{figure}
\includegraphics[width=3.25in,angle=0]{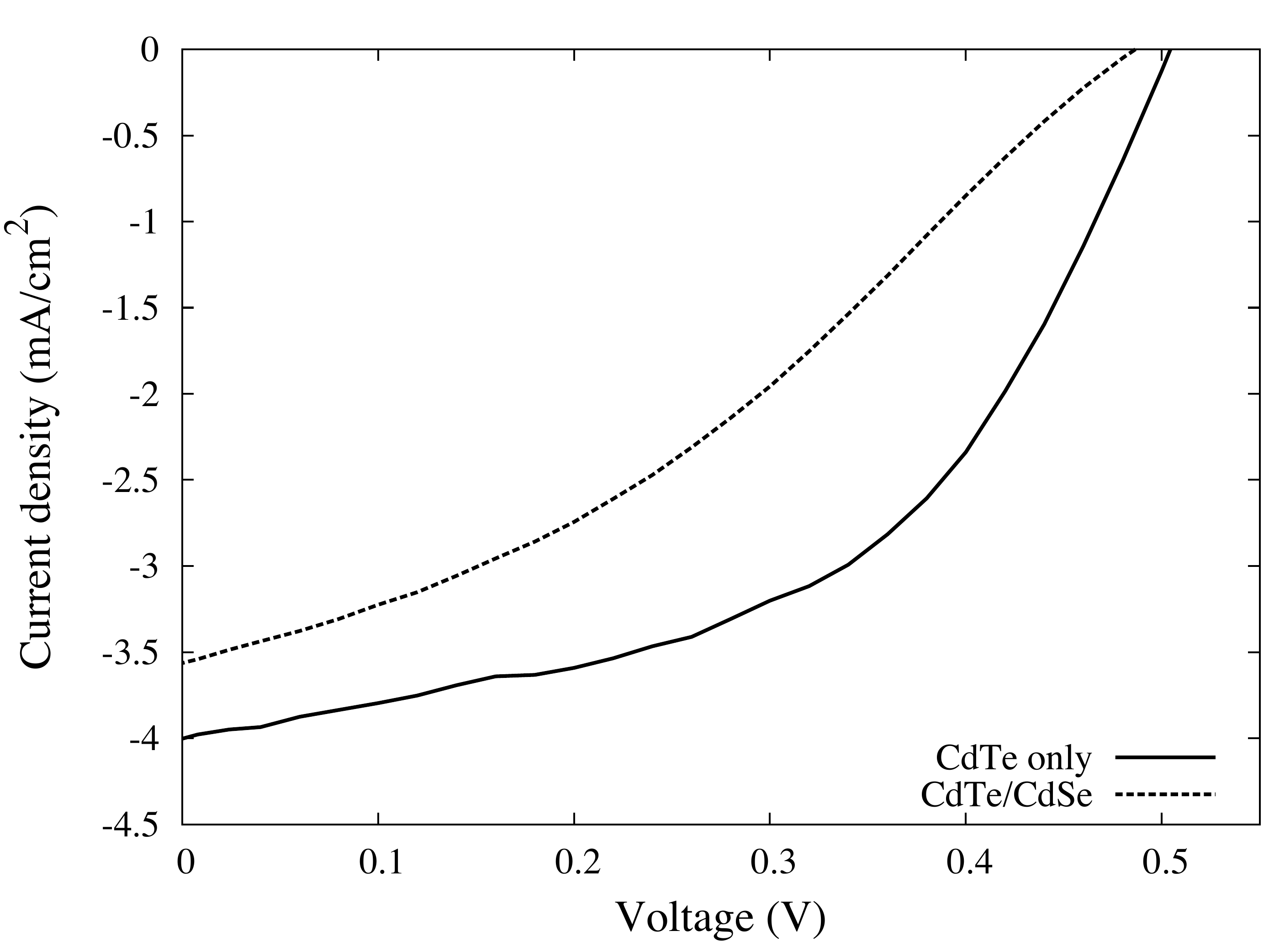}
\caption{J-V curve for CdTe-only and CdTe/CdSe devices under illumination with similar thicknesses of 420 nm. For CdTe-only device: $\rm J_{sc}$ is 4.1 $\rm mA/cm^2$, $\rm V_{oc}$ is 0.50V and fill factor is 51\%; total efficiency is 1.1\%. For CdTe/CdSe: $\rm J_{sc}$ is 3.6 $\rm mA/cm^2$, $\rm V_{oc}$ is 0.48V and fill factor is 35\%; total efficiency is 0.6\%.}
\label{IV2}
\end{figure}

The similarities (Fig. \ref {IV2}) between the CdTe-only and CdTe/CdSe devices' I-V curves suggest shared characteristics, such as junction type. However, there is a discrepancy between the two devices' EQE spectra (Fig. \ref{EQE}), i.e. the CdTe/CdSe device shows a much higher response at 400-600nm. The CdSe may be contributing this extra photocurrent, a case which lends itself to the heterojunction theory. However, based on the films' 400-500 nm thickness and near opacity, transport at long wavelengths may be hindered for the CdTe/CdSe device by morphology of the film\cite{vfc00}. Transport-limiting morphology introduced by the CdSe layer is also consistent with more minor differences between the devices, such as the lowered fill factor of the CdTe/CdSe device. 

A question remains as to why the CdTe-only device needs to be thicker than the optimal CdTe/CdSe device. We have consistently seen that thickening the device lowers the current; a device of 250 nm thickness has a $\rm J_{sc}$ of 19 $\rm mA/cm^2$, while a device of 400-500 nm thickness has a $\rm J_{sc}$ of 3 $\rm mA/cm^2$. The answer to this may be related to the formation of the CdTe layer with and without the CdSe. It seems that the CdSe layer allows for fewer microshunting paths in the thinner films required for high currents, even though it may introduce more trap states. More imaging of the CdTe-only device is needed to adequately answer this question.

\begin{figure}
\includegraphics[width=3.25in,angle=0]{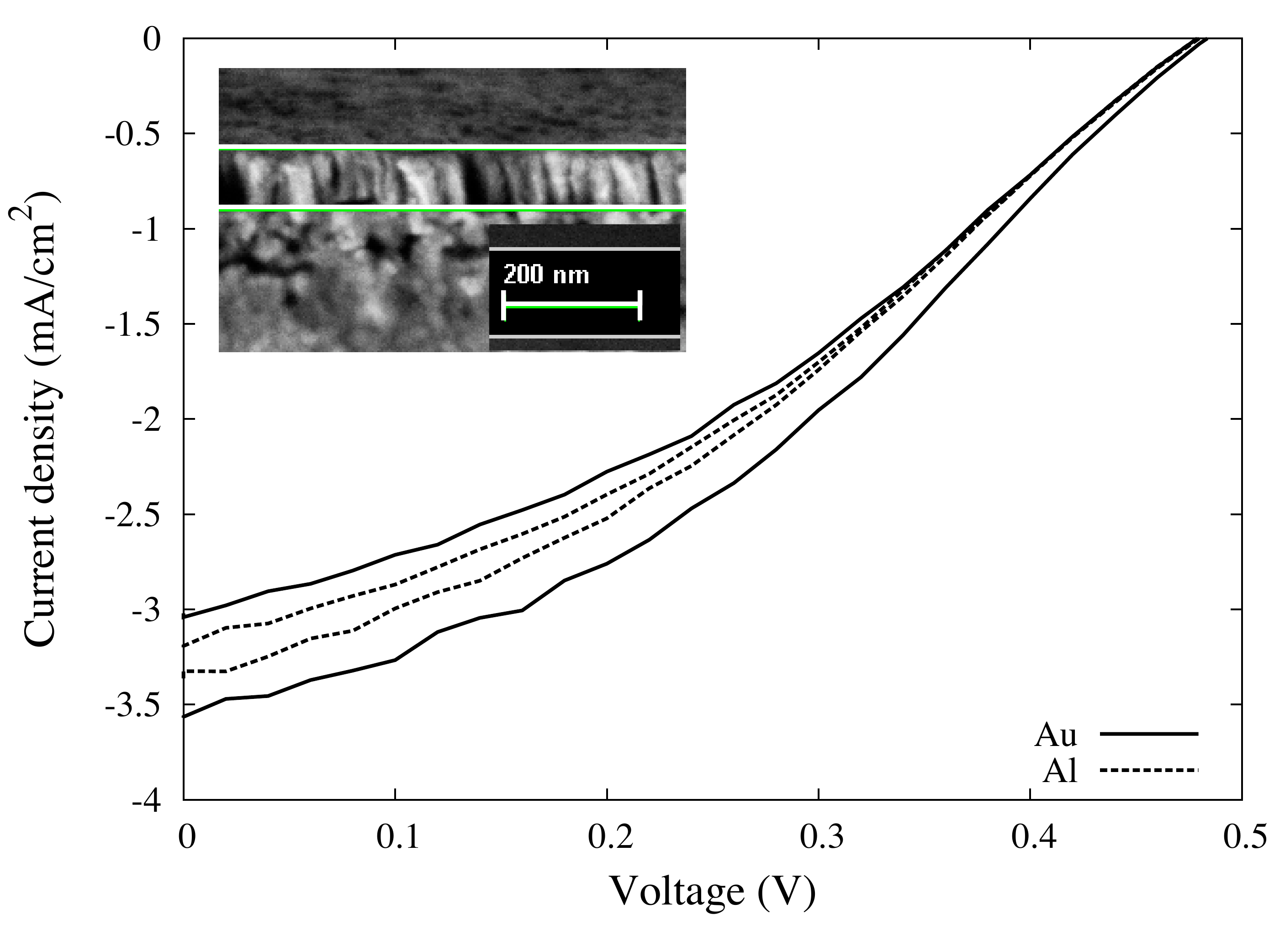}
\caption{J-V curves of CdTe-CdSe devices with both gold and aluminum electrodes. {\bf (inset)}Side view of nanoparticle film after sintering. The distance between the white cursors is 91 nm, which corresponds roughly with the CdSe layer. Below this is the CdTe layer, which does not have the same structure as the CdSe layer.}
\label{J-V}
\end{figure}

A striking feature of the nanoparticle CdTe/CdSe devices is the non-selectivity of the electrodes. Fig. \ref {J-V} shows that all-inorganic nanoparticle cells with gold and aluminum electrodes have consistently similar behavior. Ag, Ca/Al, Au and Al electrodes all yield $\rm V_{oc}$s that are statistically the same. Based on the Schottky theory of metal-semiconductor interfaces\cite {rw88}, it is unclear why this happens. Al and Ca/Al (2.87eV) have work functions less than the electron affinity of CdSe, and will form ohmic contacts with the CdSe. Ag and Au (4.26 and 4.64 eV, respectively) will form Schottky barriers; the difference between the back electrode and ITO's work functions should determine the maximum $\rm V_{oc}$. However, some electrodes such as Pd and silver paste show no appreciable $\rm V_{oc}$ at all. A thin layer of poly(ethylenedioxythiophene) (PEDOT) inserted under gold also had no $\rm V_{oc}$. These electrodes have similar work functions as gold and behave as expected for a large Schottky barrier. It is possible that the diffusive nature of certain metals contributes to the mechanism allowing for the non-selectivity of the electrodes. The PEDOT, for example, stops the gold from diffusing into the film.

The other contributing factor to the non-selectivity of electrodes is the surface states of the CdSe. SEM images of the CdTe/CdSe structure (Fig. \ref{J-V}) show that the CdSe assembles into columnar shapes after being sintered at 400C. The diffusive metals may be able to penetrate into the columns created by the CdSe, opening up enough surface states to pin the metal's work function at the level of the CdSe's conduction band.  Interestingly, preliminary work by our group on the reverse (CdSe/CdTe) structure and a CdS/nanoparticle CdTe structure exhibited normal electrode dependence\cite{o08}, a result that is under further investigation.

We have reproduced the all-inorganic nanoparticle CdTe/CdSe solar cell with efficiency up to 2.6\% without the alumina or calcium layers by increasing the thickness of CdTe. This provides a practical advantage both in cost (ALD alumina) and stability (calcium). We also have explored the possible junction types in the CdTe/CdSe device and the non-selectivity of the electrodes. The advent of a CdTe-only device and the similarities between CdTe-only and CdTe/CdSe J-V curves suggest that more work needs to be done to determine the junction type of the  CdTe/CdSe cell. Preliminary development of the CdTe-only device shows that it may have some advantages over the CdTe/CdSe device if it can be made into a thinner film.

\end{document}